\documentclass[aps,prl,twocolumn]{revtex4}

	\usepackage{graphicx}  
	\usepackage{xcolor}  
	\usepackage[colorlinks=true, linkcolor=blue, citecolor=blue, urlcolor=blue]{hyperref}
 	\usepackage{boondox-calo} % for lower-case script fonts
	\DeclareMathAlphabet{\orgcal}{OMS}{zplm}{m}{n}
		\newcommand{\M}{\orgcal{M}}
		\def\s{\mathcal s}

	\newcommand{\eq}[1]{(\ref{#1})}
	\def\be{\begin{equation}}
	\def\ee{\end{equation}}
	\def\bea{\begin{eqnarray}}
	\def\eea{\end{eqnarray}}
	\def\bean{\begin{eqnarray*}}
	\def\eean{\end{eqnarray*}}
	
	\def\bm#1{\mbox{\boldmath$#1$}}
	\def\u#1{\underline{#1}}
	
	\def\T{\textstyle}
	\def\l{\left}
	\def\r{\right}
	\def\eg{e.\,g.}

	\def\nf{n_{\!f}}
	\def\LO{{_{\mathrm{LO}}}}
	\def\NLL{{_{\mathrm{NLL}}}}
	\def\gsim{\mathrel{\rlap{\lower0.2em\hbox{$\sim$}}\raise0.2em\hbox{$>$}}}
	\def\lsim{\mathrel{\rlap{\lower0.2em\hbox{$\sim$}}\raise0.2em\hbox{$<$}}}
	\def\lg{\mathrel{\rlap{\lower0.25em\hbox{$>$}}\raise0.25em\hbox{$<$}}}
\begin{document}

\title{Low-shear QCD plasma from perturbation theory}

\author{Greg Jackson}
\email{jckgre003@myuct.ac.za}
\author{Andr\'e Peshier}
\email{Andre.Peshier@uct.ac.za}
\affiliation{
  Department of Physics, University of Cape Town, Rondebosch 7700, South Africa
}

\begin{abstract}
  We argue that the phenomenologically inferred ratio of shear viscosity to entropy density of the quark-gluon plasma, $\eta/\s \lsim 0.5$ near the deconfinement temperature $T_c$, can be understood from perturbative QCD.
  To rebut the widespread, opposite view we first show that, and why, the existing leading order result in (fixed) coupling should not be further expanded in logarithms. 
	Emphasizing then that the resummation mandatory for screening also settles the often neglected question of scale setting for the running coupling, we establish a temperature dependence of $\eta/\s$ which agrees well with constraints from hydrodynamics. 
\end{abstract}

\maketitle

RHIC and LHC experiments have provided substantial evidence that the quark-gluon plasma (QGP) behaves as an almost ideal fluid \cite{Experiment}, with an upper bound on the ratio of shear viscosity to entropy density, $\eta/\s \lsim 0.5$. While this remarkably low value clearly indicates a `strongly coupled' system, it remains a theoretical challenge to understand better {\em why} it is so low.

One popular approach to this question is via the AdS/CFT correspondence \cite{Kovtun:2004de}, which allows one to explore the strong coupling behavior of certain conformal field theories. 
Although the conjectured lower limit $\eta/\s \ge 1/(4\pi)$ from supersymmetric Yang-Mills theories does compare favorably with the observations, a rigorous connection to real-world QCD is lacking.
First attempts to compute $\eta$ by lattice QCD corroborate small values \cite{lattice}, but are hampered by the methodological difficulties of applying a static approach for a non-equilibrium phenomenon.
On the other hand, there is a widespread belief that QCD perturbation theory, as a weak-coupling method, fails to explain $\eta/\s \lsim 0.5$.
This is the perception we will scrutinize here.
	
\begin{figure}[b]
  \includegraphics[scale=.72]{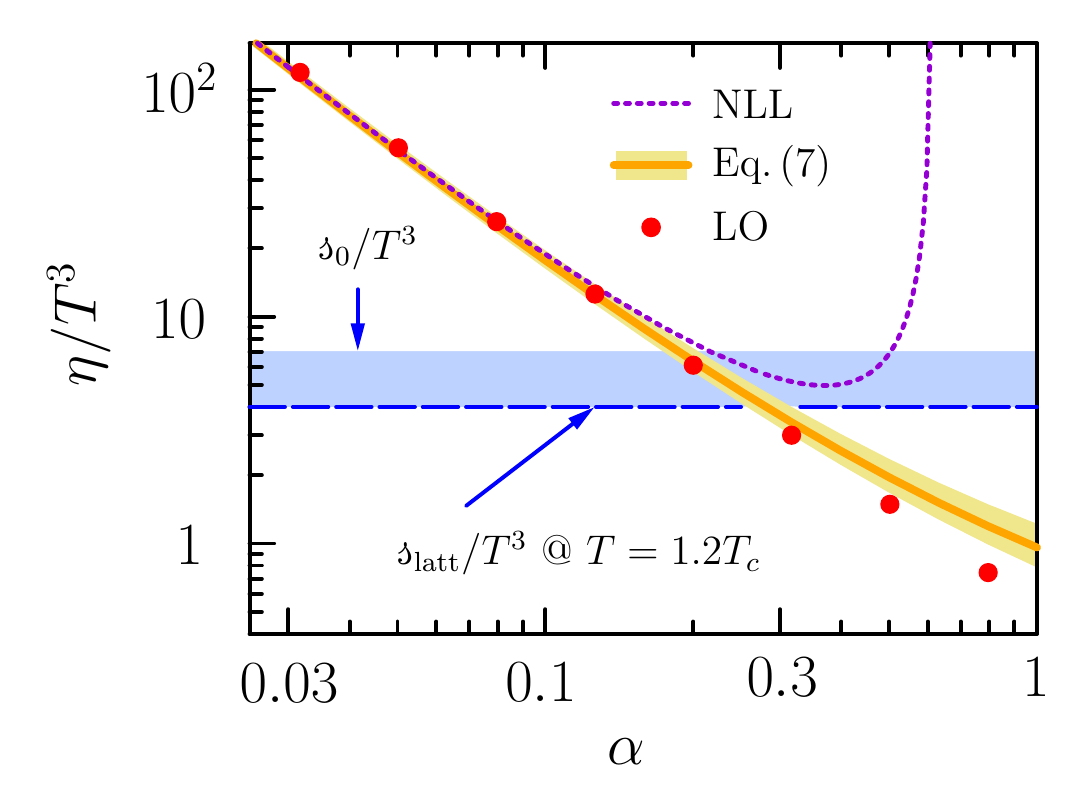}
	\vskip -3mm
	\caption{
    The viscosity, for $\nf = 0$, to LO and NLL accuracy, and from our estimate \eq{eta simple}. 
    To illustrate that $\eta_\NLL$ cannot explain $\eta/\s \lsim 0.5$ (but $\eta_\LO$ may), we also show the constraint for the entropy, $4T^3 \le \s \le \s_0$ for $T > 1.2T_c$ (see main text).
  \label{fig1}}
\end{figure}

It appears to be largely based on the {\em next-to-leading log} (NLL) formula 
\be
  \eta_\NLL (\alpha)
  =
  \frac{b T^3}{ \alpha^2 \ln(c/\alpha) } \, ,
  \label{eta NLL}
\ee
where $T$ is the temperature and $\alpha$ the coupling strength. 
The coefficients $b$ and $c$ were extracted from the {\em leading order} (LO) result $\eta_\LO$ computed numerically in a QCD effective kinetic framework \cite{AMY}. 
In the quenched limit ($\nf = 0$ quark flavors), the case we will consider mostly for argument's sake, $b \approx 0.34$ and $c \approx 0.61$. 
On general grounds, the viscosity should decrease for stronger interactions (that equilibrate velocity gradients more rapidly), which is described by \eq{eta NLL} only for $\alpha < \u{\alpha} = c/\sqrt{e}$ ($e$ is Euler's number), at which point $\eta_\NLL(\alpha)$ has a minimum. 
Numerically, ${\rm Min} \l[ \eta_\NLL \r] = 2b e T^3 / c^2$ turns out to be close to the free entropy $\s_0 = (16 + \frac{21}2 \nf) \frac{4\pi^2}{90} T^3$, see Fig.~\ref{fig1}.
Thus, since near the deconfinement temperature $T_c$ the entropy of the interacting QGP is notably smaller than $\s_0$, \eq{eta NLL} is indeed incompatible with the quite conservative bound $\eta/\s \lsim 0.5$. 

To weigh up this fact, we should see the minimum of $\eta_\NLL(\alpha)$ as a precursor to its singularity at $\alpha = c$ (marking the ultimate break-down of the NLL approximation) -- which an elementary consideration will reveal to be unphysical: In kinetic theory we may estimate \cite{Reif:1964}
\be
  \eta \approx \T\frac13 n \bar p \lambda
  \label{eta estimate}
\ee
from the density $n$ of particles that can transport a typical momentum $\bar p$ over a distance $\lambda$.
For binary interactions of relativistic particles $\lambda = (n\sigma_{\rm tr})^{-1}$, where $\sigma_{\rm tr}(s) = \int_{-s}^0 dt\, (\frac12|t|/s)\, d\sigma/dt$ is the transport cross section in terms of Mandelstam variables.
Although the `transport weight' $\frac12|t|/s = 1-\cos\theta$ suppresses the influence of the small-angle scatterings that prevail in gauge theories, $\sigma_{\rm tr}$ would still diverge logarithmically at tree-level due to the $t$-channel gluon exchange term in $d\sigma^{\rm tree}/dt \propto \alpha^2 [-us/t^2-ts/u^2-ut/s^2+3]/s^2$.
Since this would imply zero viscosity for, notably, any value of the coupling, it is a {\em necessity} to go beyond the tree-level approximation. 
In a hot QGP, the exchanged gluon acquires a self-energy of the order $\mu^2 \sim \alpha T^2$ and is thus screened, schematically $d\sigma^{\rm scr}/dt \sim \alpha^2/(t-\mu^2)^2$ for small $t$.
The typical invariant energy $s \sim T^2$ is much larger than $\mu^2$ for $\alpha \ll 1$, thus screening can be mimicked by a simple cut-off imposed on $d\sigma^{\rm tree}/dt$,
\be
  \sigma_{\rm tr}^{\rm scr}
  \to
  \sigma_{\rm tr}^{\rm cut}
  \sim
  \int_{-T^2}^{-\mu^2}\!\! dt\, \frac{|t|}{T^2}\, \frac{\alpha^2}{t^2}
  =
  \frac{\alpha^2}{T^2} \ln\alpha^{-1} + O(\alpha^2) \, .
  \label{sigma cut}
\ee
This reproduces [with $\bar p \sim T$ in \eq{eta estimate}] the parametric $\alpha$-dependence of \eq{eta NLL}, but also shows that the singularity of $\eta_\NLL(\alpha)$ is related to coinciding integration bounds in \eq{sigma cut}. 
Thus the reason why $\eta_\NLL$ cannot be extrapolated to larger $\alpha$ has to do with kinematic simplifications that become illegitimate, rather than an `breakdown' of perturbative QCD {\em per se} at $\alpha \simeq c$.

To validate this insight beyond the scope of \eq{eta estimate}, the viscosity has to be calculated from the energy-momentum tensor of the particle distribution $f(\bm p, \bm x, t)$ governed by the Boltzmann equation, $(\partial_t + \bm v \bm\nabla)f = C[f]$, when set up for the case of a collective small-gradient flow $\bm u$ that drives $f$ slightly out of local equilibrium. 
As detailed in Refs.\ \cite{AMY, Baym:1990uj}, $\eta$ can be obtained by extremizing a functional constructed from the collision term $C[f]$.
The gist of this somewhat technical calculation is \cite{JP}
\be
	\frac\eta{T^3}
	\simeq
	\Big[\, \int_0^\infty\!\! ds\, sP(s)
		\int_{-s}^0\! dt\, \frac{|t|}{2s}\, \frac{d \sigma}{d t} \ \Big]^{-1}
	+ \ldots \, ,
	\label{eta with P(s)}
\ee
if $d\sigma/dt$ (as a kernel in $C[f]$) depends only on the Mandelstam variables, and omitting terms sub-leading to the dominant small-angle binary scattering contributions.
As an aside, with $\sigma_{\rm tr}$ factorized from a positive weight $P(s)$ (that depends on how the system departs from equilibrium, see later), the convolution \eq{eta with P(s)} specifies more rigorously the `typical' momentum $\bar p$ in the elementary {\sl Ansatz} \eq{eta estimate}. 
Calculated with a {\em screened} cross section $d\sigma/dt$, \eq{eta with P(s)} resum powers of both $1/(\ln\alpha^{-1})$ and $\alpha$ -- as does $\eta_\LO$. The inverse-log expansion of $\eta_\LO$ was shown in \cite{AMY} to have zero radius of convergence.
We show here that the expansion in $\alpha$ is also ill-defined. To that end, we defer QCD particularities and argue on the basis of \eq{eta with P(s)}\footnote{%
	 even though its assumption is somewhat too restrictive for QCD, which limits its agreement with $\eta_\LO$ to leading log (LL) accuracy.}
applied to the simple model $d\sigma^{\rm scr}/dt$ which, now with correct kinematic limits, amends \eq{sigma cut} to
\be
	\sigma_{\rm tr}^{\rm scr}(s)
	\propto
	\int_{-s}^0 dt\, \frac{|t|}s \, \frac{\alpha^2}{(t-\mu^2)^2}
	=
	\frac{\alpha^2}s g(a) \, .
	\label{sigma_tr}
\ee
Here $g(a) = \ln\frac{1+a}a - 1/(1+a)$ is a monotonously decreasing, positive function of $a = \mu^2/s \propto \alpha$. 
By contrast, its `NLL' approximation, $g = \ln a^{-1} - 1 + O(a)$, becomes obviously unphysical for $a > 1/e$, leading to the same issues as seen in \eq{eta NLL} and \eq{sigma cut}.
We note first that this problem cannot be cured by higher order terms in the expansion due to the convergence radius, $a = 1$, set by the pole at $t = \mu^2$ (off the physical sheet) in $d\sigma^{\rm scr}/dt$. 
This feature of a finite radius of convergence will carry over to QCD. 
What is more, expanding $\sigma_{\rm tr}^{\rm scr}$ in $\mu^2/s \propto \alpha$ {\em before} convoluting it in \eq{eta with P(s)} with $P(s)$ is forbidden: The coefficients of $\alpha^n$ (the negative moments of $P$) are infrared-divergent, with increasing severity, since $P(0) > 0$ because \cite{JP} %as obvious from \cite{JP}
\be
	P(s) 
	=
	\int_{12}\! f_1^{(0)} \bar{f}_1^{(0)} f_2^{(0)} \bar{f}_2^{(0)}
	\big[ \, \chi^\prime(p_1) - \chi^\prime(p_2) \, \big]^2 + O(s) \, .
	\label{P(s)}
\ee
We denote by $f_i^{(0)} = [\exp(p_i/T)-1]^{-1}$ the equilibrium distribution for incoming particle $i$ in the local rest frame, $\bar{f}_i^{(0)} = 1 + f_i^{(0)}$, and $\int_i = \int d^3 p_i /[(2\pi)^3 2p_i]$ for the phase space integrals.
(The possibility $P(0) = 0$ is excluded since $\chi(p)$, which parametrizes the solution of the Boltzmann equation in the form $f(p) = f^{(0)} \l[ 1 + \chi \bar{f}^{(0)}( p_k p_l/p^2 - \frac13 \delta_{kl} ) \nabla_k u_l \r]$, cannot be strictly linear in $p$ \cite{Heiselberg:1994vy}.)
Note also that by crossing symmetry in $d\sigma/dt$, we obtain the same contribution \eq{sigma_tr} from the dressed $u$-channel term.

Now, since even the model with $d\sigma^{\rm scr}/dt$ does not have a weak-coupling expansion that allows for extrapolation to larger $\alpha$, we cannot expect so when taking into account QCD features more accurately.
In other words: Unless $\alpha\ll c$, estimates of $\eta$ cannot be based on the NLL formula \eq{eta NLL} but require at least the {\em unexpanded} (resummed) LO result $\eta_\LO$.

As a function of the coupling parameter, $\eta_\LO(\alpha)$ is monotonously approaching zero, which provokes the question for `the' value of $\alpha$.\footnote{%
	For large enough $\alpha$, the conjectured bound $\eta/\s \ge 1/(4\pi)$ would be violated.}
Before addressing this question to back up that perturbative QCD can indeed explain $\eta/\s \lsim 0.5$, let us briefly point out that $\eta_\LO(\alpha)$ is fairly well reproduced by our approximation (\ref{eta with P(s)}-\ref{P(s)}). Without needing to discuss further details of $P(s)$ we can simply rewrite the convolution in \eq{eta with P(s)} using the mean value theorem,
\be
	\frac\eta{T^3}
	\simeq
	\frac{b}{\alpha^2 g(\bar a)} \, .
	\label{eta simple}
\ee
Here we could sidestep solving the Boltzmann equation for $\chi(p)$ and infer that $1/\bm(2\int ds P(s)\bm) = b$ (the factor 2 accounts for $t \leftrightarrow u$ crossing symmetry) since \eq{eta simple} has to reproduce \eq{eta NLL} at LL accuracy.
Furthermore, $\bar a = \mu^2/\bar s = \kappa \cdot \alpha$ could be determined from a `log moment' of $P(s)$, but we will rather adjust it to match $c$ in \eq{eta NLL}, viz.\ $\kappa \to (ce)^{-1}$. 
This effectively re-incorporates sub-dominant contributions of $t$, $u$, but also $s$-channel and inelastic scatterings that were omitted in our simple scheme. 
To quantify the uncertainty of this artifice, we vary $\kappa$ by factors $2^{\pm 1/2}$ in Fig.\ \ref{fig1}, which confirms a good agreement
of \eq{eta simple} with $\eta_\LO(\alpha)$ even for $\alpha \gsim \u{\alpha}$, where the NLL result becomes qualitatively incorrect, as discussed.

Figure \ref{fig1} also depicts the rigorous bound $s > 4T^3$ on the entropy for $T > 1.2T_c$ known from lattice calculations \cite{slQCD}, to affirm that $\eta_\NLL$ cannot explain $\eta/\s \lsim 0.5$.
On the other hand, for $\alpha$ large enough $\eta_\LO$ could be compatible with $\eta/\s \lsim 0.5$ -- which brings us back to the task of specifying $\alpha$ at a given $T$.

A common prescription in the literature is to take $\alpha$ as the running coupling
\be
	\alpha (Q^2)
	=
	\l[ \beta_0 \ln( |Q^2|/\Lambda^2) \r]^{-1} 
	\label{alpha}
\ee
(where $\beta_0 = (11 - \frac23\nf)/(4\pi)$ and $\Lambda$ is the QCD parameter) at a `typical thermal scale', usually the lowest Matsubara energy modulo a factor $\xi$ of order one, $Q_T = 2\pi T \cdot \xi$.
To then have $\eta_\LO(\alpha)/\s_{\rm latt} \lsim 0.5$ at, \eg, $T = 1.2T_c$ would require $\alpha \gsim 0.4$, see Fig.~\ref{fig1}. While the resulting $\xi \lsim 0.5\Lambda/T_c$ would be $\sim 1$, quantifying the coupling (and thus the viscosity) should be based on firmer grounds.

This loose end (of having to specify the coupling {\em a posteriori}) arises because in Ref.\ \cite{AMY} $\alpha$ is treated as if it was constant. 
{\em Imposing} then $Q_T$ as the relevant scale seems counterintuitive given the importance of a whole range of momenta, parametrically $[\mu,T]$.
Rather, as put forward early \cite{Cutler:1977qm} but rarely taken into account in finite-$T$ QCD phenomenology, the relevant scale of the running coupling in, say, $t$-channel scattering should be $t$.\footnote{%
	Choosing a different scale $Q^2$ gives, by RG-invariance, correction terms  $\alpha(Q^2)\log(Q^2/t)$ which are of higher order in $\alpha(\cdot)$ but can be large.}
This rectifies \eq{sigma cut} to
\[
  \sigma_{\rm tr}^{\rm cut}
  \sim
  \int_{-T^2}^{-\mu^2}\!\! dt\, \frac{|t|}{T^2}\, \frac{\alpha^2(t)}{t^2}
  \,=\,
  \frac{\alpha(\mu^2) \alpha(T^2)}{T^2} \ln\frac{T^2}{\mu^2} \, ,
\]
hence the overall factor $\alpha^{-2}$ in \eq{eta NLL} is to be understood as a geometric mean of the running coupling at $T \sim Q_T$ and at the soft screening scale $\mu$.

To consolidate this as our second key point: Running of the coupling emerges from vacuum fluctuations, which are inseparable from thermal fluctuations. Thus for observables that require thermal screening, like the viscosity, the `scale setting' for $\alpha(Q^2)$ is unambiguous.
For this coupling renormalization, several types of radiative corrections are needed -- of which, however, only the gluon self-energy $\Pi = \Pi^{\rm vac} + \Pi^T$ contributes in Coulomb gauge due to its Abelian-like Ward identities \cite{Grozin:2008yd}.

This noteworthy feature simplifies our argument. 
Although rarely used for vacuum QCD, in Coulomb gauge it is evident that dressing \eg\ a $t$-channel Born amplitude $\sim \alpha/t$ with $\Pi^{\rm vac}(Q) = \alpha \beta_0\big[ \epsilon^{-1} + \ln( -Q^2/L^2 ) \big] Q^2$ (in dimensional regularization with scale $L$, and $Q^2=t$) gives the renormalized $\M^{\rm vac} \sim \alpha(t)/t$ with, indeed, the coupling \eq{alpha} at the scale $t$.
At $T>0$ (where Coulomb gauge is customary for other reasons), the self-energy receives the finite contribution $\Pi^T = \alpha\,\vartheta$, where the function $\vartheta \sim T^2$ depends on $q_0$ and $q$. 
Then the renormalized amplitude becomes $\M \sim \alpha(Q^2)/\bm(Q^2-\alpha(Q^2)\,\vartheta\bm)$ \cite{Peshier:2006hi}, where we emphasize that $Q^2$ also emerges as the scale for the coupling in the thermal self-energy.
This dependence of the running coupling on the virtuality carries over to the other scattering channels and then to $d\sigma/dt \sim |\sum \M_i|^2$.
Juxtapose this consistent renormalization with the common (fixed-$\alpha$) procedure: There the vacuum part in the self-energy is dropped, to give $\M^{\rm fix} \sim \alpha / (Q^2-\alpha\,\vartheta)$ with the value of the bare coupling $\alpha$ left unspecified.

This analysis allows us to easily re-instate running in the fixed-coupling calculation \cite{AMY}, where the infrared sensitive terms in $d\sigma^{\rm tree}/dt$ were screened with {\em hard thermal loop} (HTL) insertions, replacing \eg
\be
	\alpha^2\, \frac{-us}{t^2}
	\to
	\l| \alpha D_{\mu\nu}^\star(Q) Y^{\mu\nu} \r|^2 + \T\frac14\alpha^2 \, .
	\label{AMY screening}
\ee
Here $Y^{\mu\nu} = (P_1-\frac12 Q)^\mu(P_2+\frac12 Q)^\nu$, and $D^\star_{\mu\nu} = (D_0^{-1} - \Pi_\star^T)_{\mu\nu}^{-1}$ is the Coulomb HTL propagator.
The matrix element $\alpha D^\star$, which corresponds to $\M^{\rm fix}$, separates into transverse and longitudinal contributions ($i = \{t,\ell\}$), with $D_i^\star = 1/(Q^2 - \alpha \vartheta_i^\star)$.
Promoting now $\alpha$ to be $Q^2$-dependent restores the vacuum contribution and gives the {\em renormalized} amplitude
\be
  \alpha D_i^\star(Q)
	\to
  \frac{\alpha(Q^2)}{Q^2 - \alpha(Q^2)\, \vartheta_i^\star } \, .
  \label{JP screening}
\ee
The same goes for the $\alpha^2 (-ts)/u^2$ contribution in $d\sigma^{\rm tree}/dt$, with $Q^2 \to u$ in \eq{JP screening}.
It remains to discuss the terms $\alpha^2 (3-ut/s^2)$ in $d\sigma^{\rm tree}/dt$ and $\alpha^2/4$ in \eq{AMY screening}, which only give sub-leading (finite) contributions to $\sigma_{\rm tr}$ even without thermal screening.
Accordingly, the scale for the running coupling in these terms is  irrelevant for us; we set it to $(stu)^{1/3}$.
On par is the effect of inelastic scatterings -- which we neglect altogether as they affect $\eta_\LO$ by merely a few percent \cite{AMY}.
We note that although the running coupling \eq{alpha} becomes unphysical in the far-infrared domain, $|Q^2| \lsim \Lambda^2$, this effect is rendered unimportant by thermal screening $\sim \vartheta^\star_i$ \cite{JP}.

\begin{figure*}[ht]
	\includegraphics[scale=0.72]{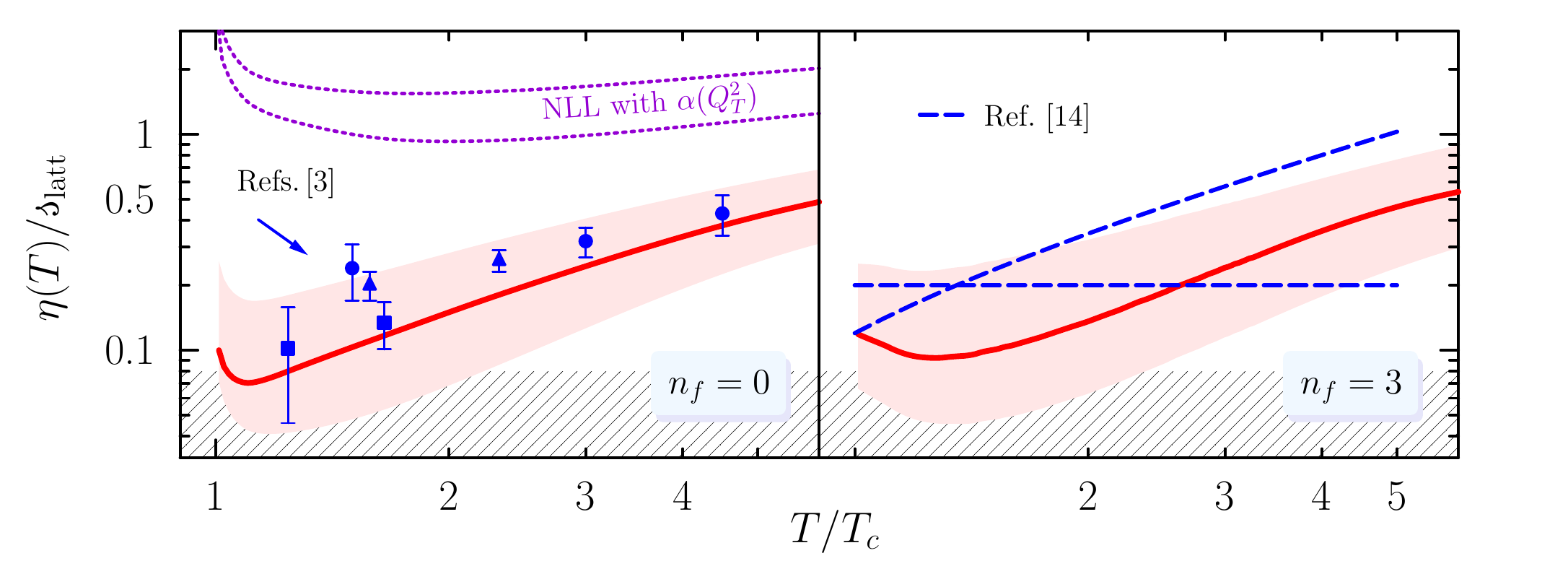}
	\vskip -3mm
	\caption{
    The viscosity in units of the interacting entropy \cite{slQCD}; the full lines show our resummed result with running coupling, the bands give the uncertainty from varying $t^\star \in [\frac12,2]T^2$, see text.
    The left panel, for the quenched limit, shows also existing lattice results \cite{lattice}, and by the dotted lines the NLL result \eq{eta NLL} varying the often imposed scale $Q_T = 2\pi T$ in running coupling by a factor of two.
	Overlayed on the right, for the physical case ($\nf = 3$), are 
    the two permissible (out of the five tested) scenarios from hydrodynamics \cite{hydro}. Hatched region: $\eta/\s \leq 1/(4\pi)$.
  }
  \label{fig2}
\end{figure*}

The HTL screening in (\ref{AMY screening}, \ref{JP screening}) is justified only for soft momenta $|Q^2| \lsim T^2$ (which is sufficient for LO accuracy). Adapting the Braaten-Yuan method \cite{Braaten:1991dd} (as done in \cite {AMY}), we omit screening for $|Q^2|>|t^\star|$ and then vary $|t^\star| \in [\frac12, 2]T^2$ to probe the sensitivity to this class of higher order contributions. Figure \ref{fig2} shows a factor of two uncertainty of $\eta$ for relevant $T$, which justifies our simplifying assumptions on the scale setting and omitting inelastic scatterings.

Such improved estimates of the viscosity only depend on the QCD scale $\Lambda$ which is of the order of $T_c$. In light of the overbearing sensitivity of $\eta$ on $t^\star$ we set $\Lambda \to T_c$ for the viscosity shown in Fig.~\ref{fig2}, normalized by the {\em interacting} entropy from lattice QCD calculations \cite{slQCD}.
For $\nf=0$ our results are compatible with existing lattice calculations of the viscosity \cite{lattice}, which  may give some guidance despite their limitations.
Interestingly, $\eta(T)/\s(T)$ hardly changes when including quarks; apparently the increased interaction rate is compensated by the density.
Our results compare favorably to recent constraints from hydrodynamics \cite{hydro} testing the average value and the $T$-dependence of the viscosity.
A fairly mild increase in $\eta(T)/\s(T)$ reflects the QCD feature of an effective coupling which weakens logarithmically.

Figure \ref{fig2} also illustrates that the NLL formula \eq{eta NLL}, supplemented by running coupling at the scale $Q_T = 2\pi T \cdot \xi$ with $\xi \in [\frac12,2]$, overestimates $\eta/\s$ by half an order of magnitude.
We have demonstrated that this estimate is misleading for two reasons, namely due to compromising the fixed-$\alpha$ LO (resummed) result by another (log) expansion and an {\em ad hoc} choice for the value of $\alpha$.
In fact, both issues are closely related: Resummation accounts for thermal screening which results from loop corrections to tree level amplitudes -- as does running coupling.

Let us conclude with two general comments, discussing first the applicability of weak-coupling methods at `larger coupling', as often relevant for heavy-ion phenomenology. 
Perturbation theory may at best give asymptotic expansions, hence higher loop corrections are not guaranteed to improve accuracy (due to lack of convergence).
Since the optimal order is expected to {\em decrease} with the characteristic $\alpha$ \cite{ItzyksonZuber}, low-order approximations can make for useful and in fact more reliable estimates. 
In the case of $\eta$, the first (and only available) candidate is the LO result for which we have emphasized the importance of running coupling: After all, $\alpha(Q^2)$ varies most where it is large.
With our second remark we justify {\em a posteriori} the use of kinetic theory  which relies on the mean interparticle distance $\bar r \sim n^{-1/3}$ being sufficiently smaller than the transport mean free path $\lambda$ \cite{Danielewicz:1984ww}.
The latter can be calculated systematically from the gain (or loss) term of our renormalized collision operator $C$, with the result that $\lambda/\bar r$ remains larger than one (althought only by a small margin) even near $T_c$ \cite{JP}.  Apparentlty, the interactions of a few partons is sufficient to maintain local equilibrium.

Treating the vacuum and thermal parts of loop corrections on the same footing, we arrive at a consistent position regarding a long-standing question: The reckoned constraint $\eta \lsim 0.5\,\s$ for the QGP produced in heavy-ion collisions can be {\em understood} on the basis of the LO viscosity -- rather than being a genuinely non-perturbative effect.

\end{document}